\documentstyle[psfig]{l-aa}

\def\lapp{\ifmmode\stackrel{<}{_{\sim}}\else$\stackrel{<}{_{\sim}}$\fi}
\def\gapp{\ifmmode\stackrel{>}{_{\sim}}\else$\stackrel{>}{_{\sim}}$\fi}

\begin{document}

\offprints{nicastro@ifcai.pa.cnr.it}
\date{Received 1 February 1998 / Accepted 25 February 1998}
\thesaurus{03(03.13.5; 03.20.4) 13(13.07.1)}

\title{$\bf BVR_cI_c$ photometry of GRB970508 optical remnant: May-August, 1997}

\author{ V.V. Sokolov\inst{1}, A.I. Kopylov\inst{1}, S.V. Zharikov\inst{1},
 M. Feroci\inst{2},
 L. Nicastro\inst{3}, and
 E. Palazzi\inst{3} }

\institute{
 {Special Astrophysical Observatory of RAS,
Karachai-Cherkessia, Nizhnij Arkhyz, 357147 Russia;
sokolov,akop,zhar@sao.ru}
\and
 {Istituto di Astrofisica Spaziale, CNR, Via Fosso del Cavaliere,
 I-00131 Roma, Italy}
\and
 {Istituto Tecnologie e Studio Radiazioni Extraterrestri CNR,
 Via P. Gobetti 101, I-40129 Bologna, Italy}
}
\maketitle
\markboth
{V. Sokolov, et al.: $ BVR_cI_c $ photometry of GRB970508 optical remnant.}
{V. Sokolov, et al.: $ BVR_cI_c $ photometry of GRB970508 optical remnant.}

\begin{abstract}
We present the results of photometric observations
of the variable optical source associated to the remnant  of
the  gamma-ray  burst GRB970508 performed at the Special Astrophysical
Observatory of the Russian Academy of Science (SAO RAS) from 
May to August 1997. The observations were carried out with the standard  
(Johnson-Kron-Cousins) photometric $ BVR_{c}I_{c}$ system using the  
1-meter  and 6-meter telescopes. 
The brightness of the  optical  source  increased 
from $R_c=21.19\pm0.25$ (May  9.75  UT) to $R_c = 19.70\pm0.03$  (May 
10.77 UT), whereupon it was decaying in all the four $BVR_{c}I_{c}$ 
bands during about one month after the burst. Between the 3rd  and  the 
6th night after the GRB970508 event, the flux decrease of the 
optical source follows an exponential law in all the 
four bands. In that period the broad-band spectrum  of  the  object 
does not change and can be approximately described by the power  law 
$F_{\nu}\propto\nu^{ -1.1}$.  
The  subsequent  observations in $BVR_{c}I_{c}$ bands have shown 
a reduction of the source brightness decrease rate. 
The source flux decay after the maximum in the $ R_c $ band is well
described by a single power-law $ F \propto t^{-1.171(\pm0.012)}$ 
for the whole set of observations during 86 days.  
In the other bands the decay seems to slow down from the 31th day onwards. 
In particular, in the $ I_c $ band the source magnitude is about 23.1 from the
36th day after the GRB to the end of the observations (August 4, 1997). 

\keywords{ gamma-rays: bursters -- optical transients, CCD photometry}

\end{abstract}

\section{Introduction.}

   Since their discovery in 1969 (Klebesadel et al. 1973) 
the gamma-ray bursts (GRBs) phenomenon is the subject 
of intensive observational and theoretical studies. 
The basic problem before 1997 was the lack of reliable counterparts 
to the sources of GRB 
(in quiet or transient state) in wave bands other than gamma-rays 
and, as a consequence,  the  total uncertainty of their distance.  
The situation changed after the 
Gamma-Ray Burst of February 28, 1997, GRB970228 (Costa et al. 1997a)
detected by the Italian-Dutch satellite BeppoSAX (Boella et al. 1997).
Thank to the fast and accurate positioning of GRBs 
(a few arcminutes)
obtainable through the
combined capabilities of the Gamma-Ray Burst Monitor (GRBM) and Wide Field
Cameras (WFCs) onboard this satellite, an X-ray and optical afterglow were
observed for the very first time (Costa et al. 1997b,
van Paradijs et al., 1997).

   However, as was found out later, 
the optical data on the first GRB afterglow in the history were obtained 
with different instruments and in different photometric conditions 
and refer to different photometric systems and bands.  
The  source brightness in different filters was determined at
different times and besides, only a few  reliable  
brightness determinations (not upper limits) have been gained while 
the object was bright enough. In the end, it  led  to  the  
uncertainty  in  the interpretation of optical observations (see
for example Sahu et al. 1997, Galama et  al. 1997 and the Proceedings of the
$4^{th}$ Huntsville GRB Symposium).
But at least, one  thing  became  clear after GRB970228:  
the optical source --  a gamma-ray burst remnant (GRBR)  --  
can be observed during rather a long period, 
from one to several days, in sufficiently bright phase, 
with the brightness of $ V<23$.

   By the end 1997, the optical afterglow of GRBs 
was reliably detected in 3 cases: GRB970228, GRB970508, GRB971214.  
While the first and third optical transient sources (OTs) 
were not observed at SAO RAS close to their  maximum brightness, 
for the second OT, related to GRB970508 
reported by the BeppoSAX team (Costa et al. 1997b; Piro et al. 1998) 
essential optical data helping to elucidate common observational signs 
of optical GRBRs have been obtained at SAO with 
the 1-m and 6-m telescopes. The second GRB with an optical 
afterglow was in a very favorable place in the northern sky
(in  the  Camelopardalis  constellation).  The high declination 
of $\sim +79^\circ$ allows us to observe this object at SAO at all dates
being at a zenith distance of less than $57^\circ$. 

Preliminary results of multi-color photometry 
of the GRB970508 optical remnant were reported elsewhere 
(Sokolov et al. 1997). 
In this paper we report the results of $ BVR_cI_c $ photometry 
from May to August 1997 discussing the correlation between the features
of the optical light curves and those observed at other wavelengths. 
The second  section  deals  with  the 
search and detection of this optical source with the SAO telescopes 
and describes the $ BVR_{c}I_{c}  $  photometry 
with the 1-m and 6-m  telescopes.  
The third section is dedicated to the results of CCD photometry 
in the phase of the source brightness maximum  
and to the following brightness fading in different filters 
up to the  87th  day after the burst. 
The fourth section  is dedicated to the discussion of the results 
and to the comparison of the optical light curves with those in
X-ray and radio.

\section{GRB970508 optical counterpart search.}

The alert of a new GRB detection by BeppoSAX reached SAO only 3.5 hours after
the high energy event (May 8.904 UT).
Due to morning twilight, follow-up observations could start at the 1-m telescope
on May 9.74 UT. The $5^{\prime}$ WFC error box was
completely covered with the CCD mosaic of 29 images
in $ R_c$ band with 300 and 600 sec exposure times.
The 1-m (Zeiss-1000) telescope is equipped with a CCD photometer
ISD015A, $ 520\times580$ pixels
corresponding to the field of view $ 2.^{\prime}0\times 3.^{\prime}6$.
The images  from the 1-m telescope were compared to the corresponding fields
of the Digitized Sky Survey (DSS). 
No new bright object was found up to the DSS limit for this field.

\begin{figure*}[t]
\centerline{
   \vbox{\psfig{figure=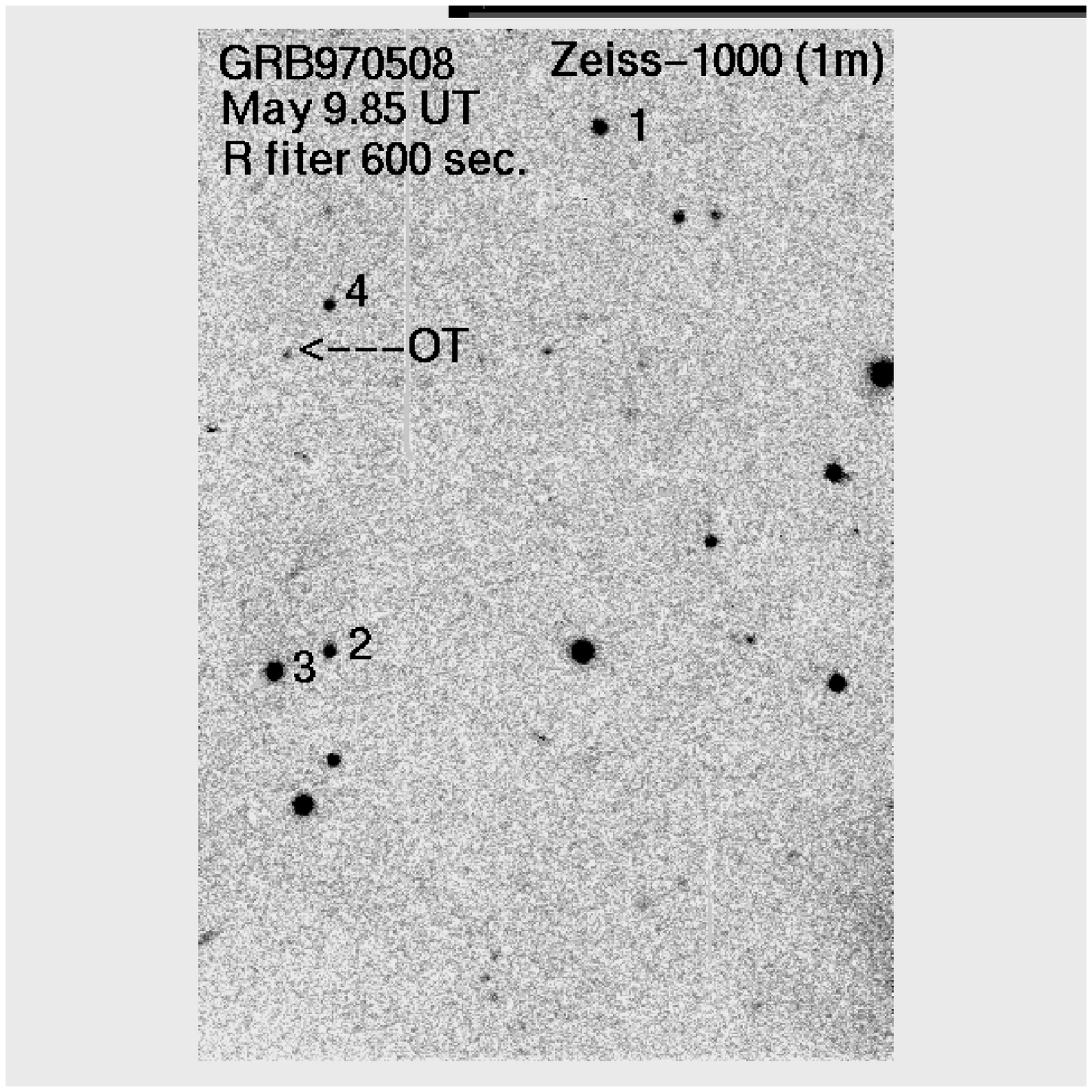,width=11cm,%
 bbllx=151pt,bblly=440pt,bburx=440pt,bbury=745pt,clip=}}
\hspace{-8.5cm}\vspace{-8cm}
   \vbox{\psfig{figure=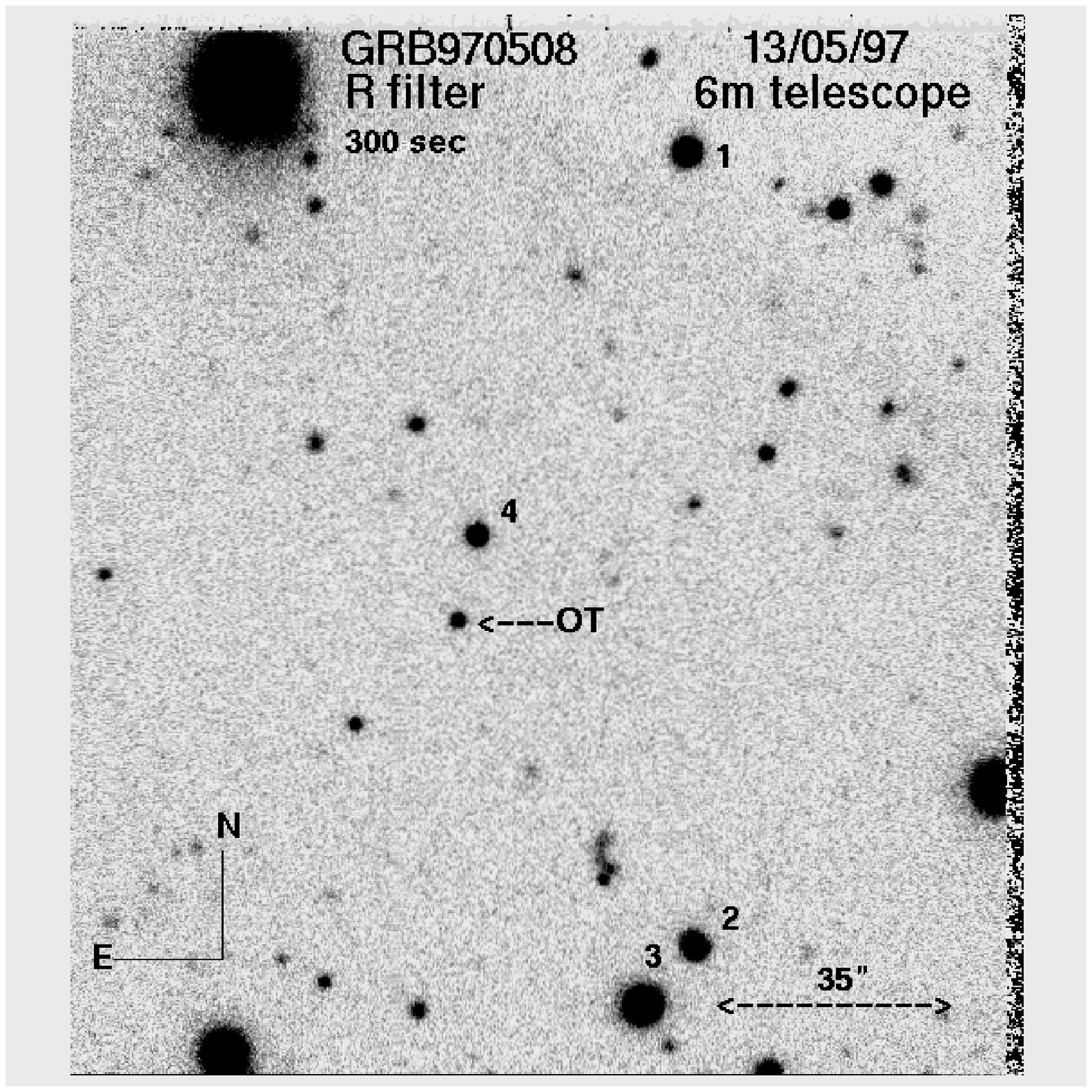,width=8cm,%
 bbllx=95pt,bblly=305pt,bburx=485pt,bbury=750pt,clip=}}
\vspace{8cm}
}
\caption{ The field near the optical source of GRB970508 imaged  with 
the  1-m  telescope  (before  the  maximum   of   the   transient 
brightness) and with the 6-m telescope (the 4th night  after  the 
maximum) in $ R_{c}$ band.  The  object  brightness  in  both  images 
corresponds  to  $  R_{c}\approx  21.1$. The stars -- secondary 
photometric standards -- are indicated.}
\end{figure*}

On the next night, May 10/11, a refined WFC error circle 
position was available:
$ \alpha_{2000}=06^h53^m28^s $; $ \delta_{2000}=+79^o17^{\prime}.4$ with
a $3^{\prime}$ error radius (99\% confidence level).
Photometric observations of GRB970508 field were then continued with the 6-m
telescope. The $ 1040\times1160$ pixel CCD chip ``Electron ISD017A'' installed
at the Primary Focus has a field of view of
$ 2.^{\prime}38\times2.^{\prime}66$.
A $ 2\times2$ binning mode was employed,
so that each of the
$ 520\times580$ zoomed pixels has angular size of
$ 0.^{\prime\prime}274 \times 0.^{\prime\prime}274$.
The gain is  $ 2.3 e^-$ per DN (Data Number).
The readout noise is about $ 10e^-$.

The first image at the 6-m telescope was obtained on May 10.76 UT and
a variable object was discovered by comparison with the image taken the night
before.
Its brightness, from May 9.85 UT to May 10.76 UT, increased
of about 1.5 magnitudes. This object was first reported by Bond (1997)
as a possible optical counterpart of GRB970508 and was independently
found in our data only about 0.5 day later.
The summary of the observations carryed out at SAO during
the first week after the GRB is reported in Table 1. The exposure times in seconds
are shown for each filters.

\begin{table}[t]
\begin{center}
\caption{First week observations of GRB970508 optical remnant.}
\begin{tabular}{ccccccc} \hline
  night &   UT     &  telescope &      $ B $ &    $ V $ &   $ R_{c}$&     $ I_{c}$ \\
        &   (May)   &           &      (s)    &    (s)      & (s)    &  (s)    \\ \hline
  09/10 &   09.75   &     1-m    &            &          &     300   &              \\
   -"-  &   09.85   &     -"-    &            &          &     600   &              \\
  10/11 &   10.77   &     6-m    &       300  &     200  &     100   &        300   \\
   -"-  &   10.93    &    -"-     &      300   &    200   &    100    &       300   \\
  11/12 &   11.76    &    -"-     &      450   &    300   &    150    &       450   \\
  12/13 &   12.87   &     -"-    &       450  &     600  &     150   &        900   \\
  13/14 &   13.88   &     -"-    &      1200  &     600  &     450   &        450   \\ \hline
\end{tabular}
\end{center}
\end{table}

\begin{table}[th]
\begin{center}
\caption{ Secondary photometric standards in the OT GRB970508 field. }
\begin{tabular}{ccccccc} \hline
 \# &   $\alpha_{2000}$&  $\delta_{2000}$ &   $ B$   &  $ V$   & $ R_c$ & $ I_c$\\ \hline
  1 &  06:53:37.19     & 79:17:30.7            &   20.44  &   19.14 &  18.31 &  17.53 \\
  2 &  06:53:36.30     & 79:15:30.0            &   19.93  &   19.17 &  18.71 &  18.27 \\
  3 &  06:53:39.23     & 79:15:21.1            &   17.94  &   17.40 &  17.06 &  16.71 \\
  4 &  06:53:48.50     & 79:16:32.7            &   21.93  &   20.43 &  19.49 &  18.53 \\ \hline
\end{tabular}
\end{center}
\end{table}

Observations were carried out with filters closely matching
the $ BVR_{c}I_{c}$ Johnson-Kron-Cousins system.
The data were processed using the ESO-MIDAS software.
Standard data reduction includes bias subtraction,
flat-fielding and cosmic particle traces removal.
Photometric conditions 
remained stable during two nights of May 13/14  and May 21/22.

Four bright stars (Fig. 1) in the GRB970508 field were used as
secondary photometric standards.
Magnitudes of these stars
were determined on the night of May 13/14 with good photometric
conditions using four standard stars in the field of
PG1657+078 (Landolt, 1992).
Zero-point errors are better than $ 0.05^m$.
Coordinates and magnitudes of secondary photometric standards
are  given in Table 2. Our $ R_c$ magnitudes of
stars 2, 3, 4 are $ 0.20\pm0.01$ higher than the magnitudes measured
by Schaefer et al. (1997). 
The application of the analogous photometric procedure in 
$ BVR_{c}I_{c}$ Johnson-Kron-Cousins system, used at SAO RAS, 
is given in details in the paper by 
Kurt  et  al. (1998).

\begin{figure*}[t]
\centerline{
   \vbox{\psfig{figure=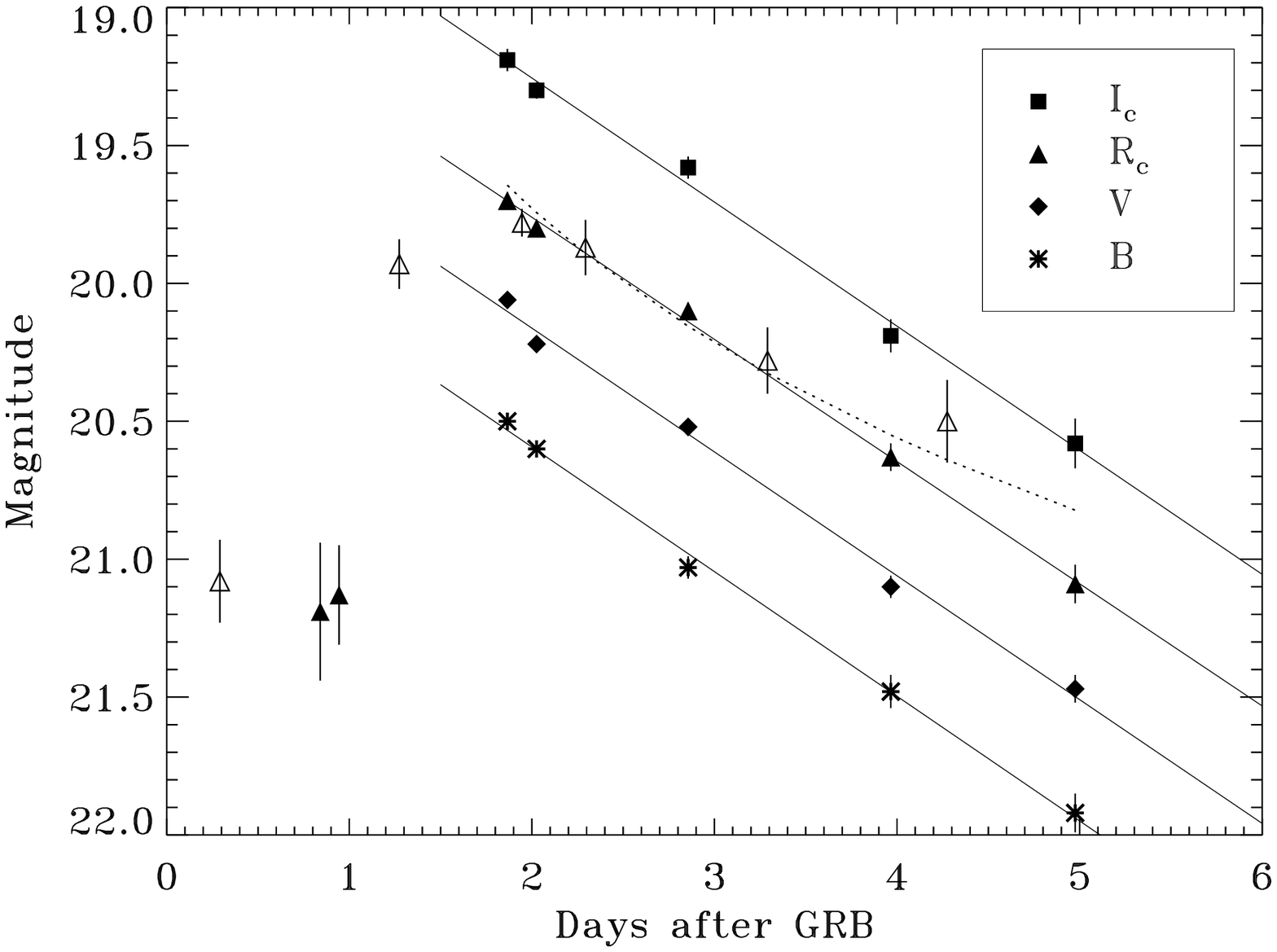,width=13cm,%
 bbllx=20pt,bblly=300pt,bburx=570pt,bbury=730pt,clip=}}
}
\caption{ The light curves of GRB970508 optical counterpart during
5 days after the burst. SAO RAS (filled symbols), 
Loiano (Mignoli M. et al. 1997, $ t-t_0 = 1.95$) and Palomar 
(Djorgovski et al. 1997, transformed to $ R_{c} = r - 0.34 + A_{r}$)
(open triangles) magnitudes with their errors are shown.
Lines correspond to equations (1) of exponential decline of brightness
reported in the text. For $R_c$ also the best fit power law is shown
(dotted line).
}
\end{figure*}
\begin{figure*}[t]
\centerline{
   \vbox{\psfig{figure=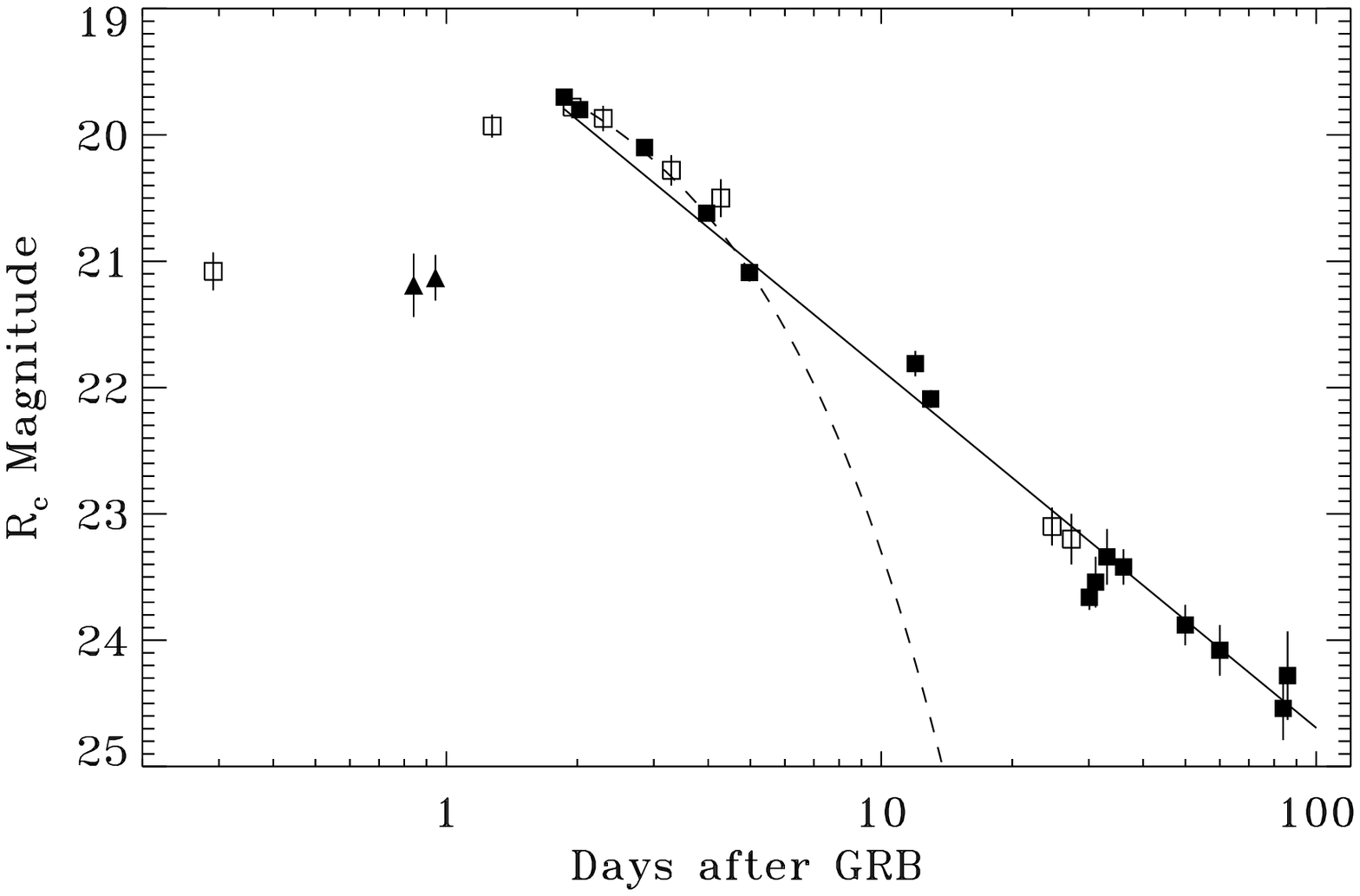,width=13cm,%
 bbllx=30pt,bblly=325pt,bburx=580pt,bbury=675pt,clip=}}
}
\caption{ $ R_c$ light curve of the optical counterpart of GRB970508 during
$\approx 86$ days after the burst.
SAO RAS (filled marks) and 
Palomar (Djorgovski et al. 1997), Loiano (Mignoli et al. 1997),
 HST (Fruchter et al. 1997),
Keck II (Metzger et al. 1997)  (open squares,
transformed from Schaefer's photometric system to ours)
 magnitudes are shown.
Lines correspond to power (up to $t-t_0=83.940$) and exponential law (dashed)  for fading brightness.
}
\end{figure*}

\section{The brightness maximum phase and following source fading}

We define as {\em source brightness maximum phase} the  period 
when the optical object brightness increased to the maximum (or close to the 
maximum) and then began to drop down to 
$V\approx  21.5 $  and $ R_{c}\approx  21.1$.   
It corresponds approximately to the first week of observations 
after the GRB when the optical source  could 
still be observed with many telescopes  -  big  and  medium  ones. 
The results of the observations performed during these nights
at SAO together with those from Palomar and Loiano are
shown in Fig. 2.
Many more observations other than those from Palomar and Loiano were performed,
but the different photometric systems used and relatively smaller accuracy
of the measurements do not allow a better evaluation of the OT behavior.

   Johnson-Kron-Cousins magnitudes in the $ BVR_{c}I_{c}$ bands
with the associated errors for GRB970508 optical counterpart are given
in Table 3. Given the BeppoSAX/GRBM trigger time $t_0$ = May 8.904 UT,
we note (see Fig.2): \\
 1) from $ t-t_0 = 0.226$ to $ t-t_0 = 0.944$ 
the  $ R_c$ brightness of the object seems to remain constant
(Castro-Tirado et al. 1997;  Djorgovski et al. 1997); \\
 2) from $ t-t_0 = 0.944$  to $ t-t_0 = 1.866$  the $R_c$ brightness increased 
of 1.5 mag. The magnitude increase rate using our 1-m data and 
the data from Palomar (Djorgovski et al. 1997) 
was about 0.12 mag per hour; \\
 3) at $ t-t_0 = 1.866$ the $ R_c$ magnitude reached 
the brightness maximum of 19.70 and from that date 
a decline of brightness began; \\ 
 4) the magnitude decrease during the beginning
(2--5 days) of the fading phase follows an exponential law in all four bands:
\begin{eqnarray}
    B  = 19.689(\pm0.036)+0.452(\pm0.014)(t-t_0) \nonumber  \\
    V  = 19.264(\pm0.053)+0.449(\pm0.020)(t-t_0)   \\
    R_c= 18.874(\pm0.029)+0.443(\pm0.011)(t-t_0)\nonumber \\
    I_c = 18.355(\pm0.050)+0.450(\pm0.019)(t-t_0) \nonumber
\end{eqnarray}
 
A power law did not fit the data for the brightness fading after the maximum. 
For example for the $R_c$ data it is $\chi_{n}^2=0.97$ for the
exponential fit while it is $\chi_{n}^2=4.5$ for the power law.
All the $R_c$ data from $ t - t_0 = 1.866$ to $ t-t_0 = 4.976$ 
(10.77 UT -- 13.88 UT) are used: SAO, Loiano and Palomar.

\begin{table*}[t]
\begin{center}
\caption{GRB970508 OT photometry results}
\begin{tabular}{lrcccc} \hline
  UT & $ t-t_0$    &      $ B$   &  $ V$          &   $ R_c$      &      $ I_c$  \\ \hline
 9.75 May & 0.841  &                &               &  $21.19\pm0.25$ &                \\
 9.85     & 0.944  &                &               &  $21.13\pm0.18$ &                \\
 10.77    & 1.866  & $20.50\pm 0.03$& $20.06\pm0.03$ &  $19.70\pm0.03$ & $19.19\pm0.04$ \\
 10.93    & 2.026  & $20.60\pm 0.03$& $20.22\pm0.03$ &  $19.80\pm0.03$ & $19.30\pm0.03$ \\
 11.76    & 2.856  & $21.03\pm 0.04$& $20.52\pm0.03$ &  $20.10\pm0.03$ & $19.58\pm0.04$ \\
 12.87    & 3.966  & $21.48\pm 0.06$& $21.10\pm0.04$ &  $20.63\pm0.05$& $20.19\pm0.06$ \\
 13.88    & 4.976  & $21.92\pm 0.07$& $21.47\pm0.05$ &  $21.09\pm0.07$& $20.58\pm0.09$ \\
 20.795   & 11.891 &                & $22.10\pm 0.18$&               &                      \\
 20.875   & 11.971 &                &               &$ 21.81\pm 0.10$&                      \\
 20.970   & 12.066 &                &               &                &$ 21.27\pm  0.13$          \\
 21.892   & 12.988 & $23.12\pm 0.35$&               &$ 22.09\pm 0.07$&                      \\
 21.908   & 13.004 &                &$22.18\pm 0.14$&                &                      \\
 21.942   & 13.038 &                &               &                &$ 21.96\pm  0.13$          \\
 07.917 Jun.  & 30.085 &                &               &$ 23.66\pm 0.10$&                      \\
 08.964   & 31.060 & $24.65\pm 0.25$&               &                &                      \\
 08.980   & 31.075 &                &$23.84\pm 0.24$&                &                      \\
 08.991   & 31.087 &                &               &$ 23.54\pm 0.20$&                      \\
 10.928   & 33.024 &                &               &$ 23.34\pm 0.22$&                      \\
 10.940   & 33.036 &                &$24.08\pm 0.25$&                &                      \\
 10.945   & 33.041 &                &               &                 &  $23.09\pm 0.35$    \\
 13.954   & 36.050 & $24.62\pm 0.14$&               &                &                      \\
 13.966   & 36.061 &                &               &$ 23.42\pm 0.14$&                      \\
 13.992   & 36.088 &                &               &                &$ 23.30\pm  0.35$          \\
 27.865   & 49.961 & $24.50\pm 0.22$&               &                &                      \\
 27.873   & 49.969 &                &$24.17\pm 0.16$&                &                      \\
 27.893   & 49.989 &                &               &$ 23.88\pm 0.16$&                      \\
 27.910   & 50.006 &                &               &                &$ 23.05\pm  0.14$          \\
 07.946 Jul.  & 60.042 & $25.07\pm 0.21$&$24.62\pm 0.25$&$ 24.08\pm 0.20$&                      \\
 07.961   & 60.057 &                &               &                &$ 23.10\pm  0.15$          \\
 31.843   & 83.940 &                &               &$ 24.54\pm 0.25$&                      \\
 31.857   & 83.952 &                &$24.60\pm 0.24$&                &                      \\
 31.862   & 83.958 &                &               &                &$ 23.08\pm  0.23$          \\
 02.807 Aug.  & 85.902 &                &               &$ 24.28\pm 0.35$&                      \\
 02.825   & 85.920 &                &$24.59\pm 0.25$&                &                      \\
 02.855   & 85.950 & $25.32\pm 0.25$&               &                &                      \\
 03.989   & 87.085 &                &               &                &$ 23.04\pm  0.14$          \\   \hline
\end{tabular}
\end{center}
\end{table*}

The spectrum of the object was close to a power law and its slope
$  F_{\nu}\propto\nu^{ -1.2} $ did not change in time, with
      $$  (B-V)=0.43, \ \ (V-R_c)=0.39, \ \ (R_c-I_c)=0.52$$
Taking into account galactic absorption $  E(B-V)=0.03  $ gives
       $$  F_{\nu}\propto\nu^{ -1.1} $$  and the following color indeces:
       $$  (B-V)_0=0.40, \ \ (V-R_c)_0=0.37, \ \ (R_c-I_c)_0=0.50$$

After the 5th day the decline of brightness is slowed down.  
In the $BVR_c$ bands during about 85th day and $I_c$ band until 
the $ \approx 36$th day, the light curve can be described
by a power-law relation,
\begin{eqnarray}
    B   =  19.713(\pm0.139)+3.000(\pm0.048)\times \log(t-t_0) \nonumber \\
    V   =  19.323(\pm0.135)+2.860(\pm0.046)\times \log(t-t_0)          \\
    R_c  = 18.865(\pm0.092)+2.927(\pm0.031)\times \log(t-t_0) \nonumber \\
    I_c  = 18.313(\pm0.151)+2.961(\pm0.066)\times \log(t-t_0) \nonumber
\end{eqnarray}
with
 $$\alpha_{B}= -1.200(\pm0.019)$$
 $$\alpha_{V}= -1.144(\pm0.018)$$
 $$\alpha_{R_{c}}= -1.171(\pm0.012)$$
 $$\alpha_{I_{c}}= -1.184(\pm0.026),$$
neglecting the fact that for the first 5 days the light curve decay 
is better described by the exponential law in all the 4 bands. 
Figure 3 shows this averaged power-law light curve for the $R_c$ band 
up to about 84 days ($t-t_0=83.940$) after the burst.

Starting from the $\approx 36$th day after the burst, the object did not show
the further fall of the brightness in the $I_c$ filter up to $t-t_0=87.085$.
The magnitude of the object in the $I_c$ filter in that period was
stable around  $ I_c = 23.07$.
Figure 4a shows the light curves in the $BVR_cI_c$ bands up to day 87.085.

\section{Discussion. Comparison of optical, X-ray and radio 
brightness curves}

The data obtained with the SAO RAS 1-m and 6-m telescopes
allow us to divide the GRBR970508 brightness change curve 
into three stages (Fig. 2, 4 and Table 3): \\
1) the increase of brightness on time scale of about one day; \\
2) the exponential brightness fall during about 4 days with
a stable broadband power-law spectrum;     \\
3) the further brightness fading according to a power law.     \\
Nevertheless, from the 36th day after the burst the object did not show
the further fall of the brightness in the $I_c$ filter.
After 87 days, the deviation from the average power law (2) 
achieved already $\approx 1$ mag.  

As Figure 3 shows, our observations with the 6-m telescope alone
can be approximately described by a single law in the $R_{c}$ filter, 
but in the phase of brightness maximum (Fig. 2), 
when all fluxes were  measured  with  the smallest errors, 
we see the same deviations from the ``average" 
power law in {\bf all} 4 bands. 
Garsia et al. (1997) also mention that the brightness fading after  the 
maximum cannot be well described by a single power law. 

As discussed by Piro  et  al. (1998), after about 30 seconds from the high
energy event, the GRB970508 X-ray flux  began  to 
fall according to a power law with  a slope  of  $-1.1\pm 0.1$. 
Nevertheless, in about 16.6 hours after the GRB the flux  decay
showed a temporal behavior analogous to the optical one during the
first hours after the GRB (Pedersen et al. 1998).
(We did not observe at that time, but in the R band
the optical flux was either stable or showed a slow decrease 
in the first 8 hours after the GRB.)
The features of the optical curves immediately after the maximum
(May 10.77 UT) correlate with those of the X-ray transient light curve
(Piro  et  al. 1998).

In Figure 4 the SAO RAS $BVR_cI_c$ brightness curves are presented together
with the radio at 8.46 GHz (Frail et al. 1997) and the X-ray (BeppoSAX
NFIs, Piro et al. 1998).
Apparently, the most interesting moment in this phase 
was the sharp fall of the X-ray flux 
and its subsequent fast increase which was observed in details 
in the BeppoSAX MECS range 2--10 keV, but unfortunately no simultaneous
optical observations were performed.
The subsequent optical brightness increase which was observed also at SAO
(Fig. 2) began with a small ($ \approx 5.5$ hours) delay. (Though now 
it can be supposed that before that moment  there  was  a  genuine 
optical brightness minimum  missed  by the observers.) Piro et al. 
(1998) notice that the following behavior of X-ray  flux  deviates 
from the power law. In the X-ray range  this deviation,  lasting 
about 4 days, took about 30\% of the energy released in the ``afterglow" 
part of the power-law drop. It is about 10\% of the gamma-ray energy itself. 
During the same period the optical decay does not follow a power-law too, but
the exponential one (Fig. 2 and 3).
This is in favor of a common origin of the X-ray and optical events. 
Taking into account the ratio of X-ray to optical flux during the 4 days after
the optical maximum, more than 16\% of the energy
of the GRB event itself was released in optical together with X-rays.
Thus, it is believable  indeed  that 
the energy of the GRB remnant is not only determined by the afterglow, 
but there is also an intrinsic burst activity of the source 
which has an energy comparable to that of the burst 
in a time scale 10.000 times longer than the duration of the GRB itself. 

\begin{figure*}[t]
\centerline{
   \vbox{\psfig{figure=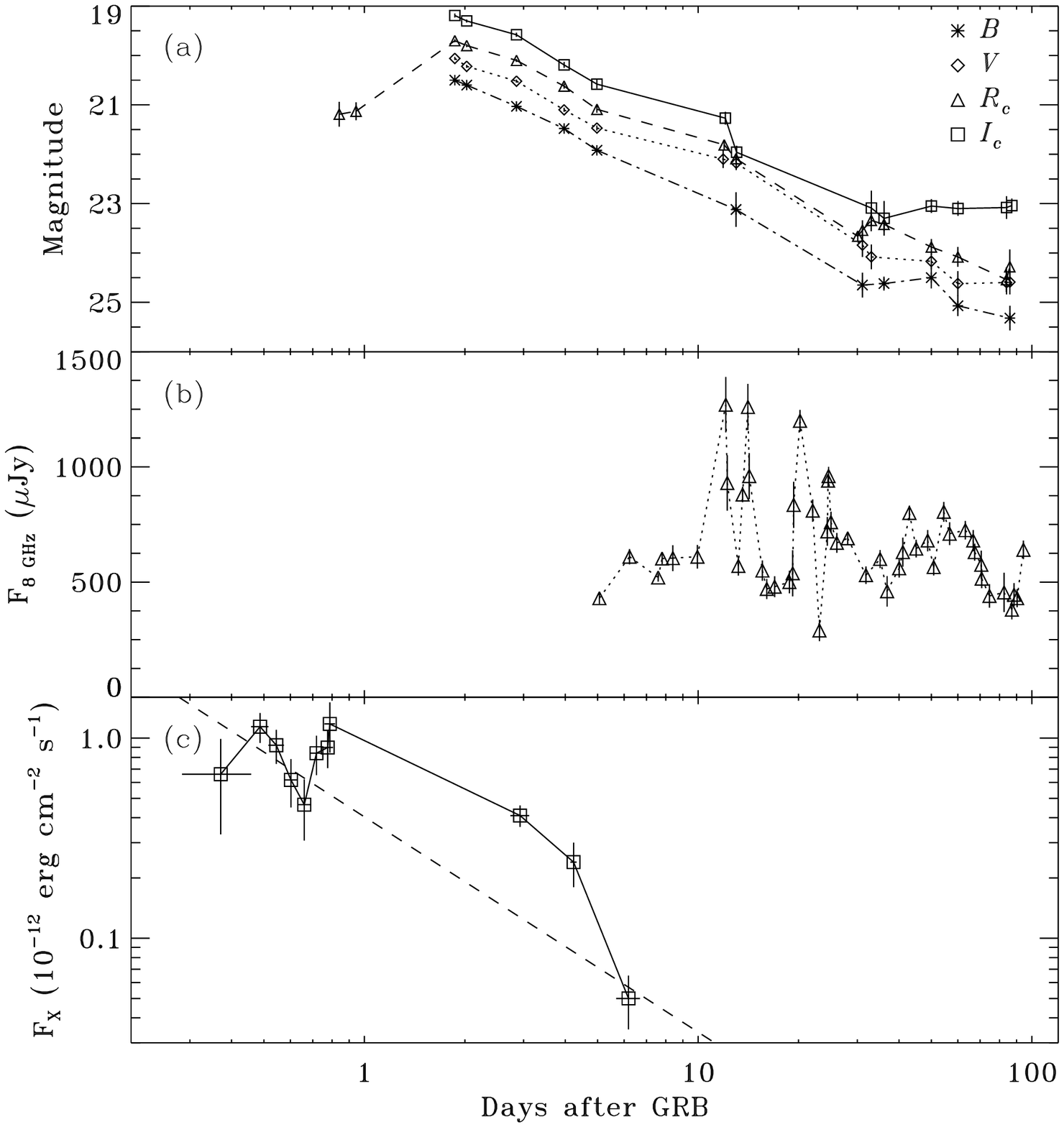,width=16.5cm,%
 bbllx=20pt,bblly=185pt,bburx=570pt,bbury=780pt,clip=}}
}
\caption{GRB970508 light curves: (a) $BVR_cI_c$ SAO RAS data;
  (b) VLA/VLBA 8.46 GHz data (Frail et al. 1997);
  (c) 2--10 keV BeppoSAX MECS X-ray data with the $t^{-1.1}$ decay law
      (Piro et al. 1998). The power law with the slope of $-1.1$ is shown
 (dashed line).}
\end{figure*}

When comparing the radio and optical behavior of the  source  
(Frail et al. 1997; Taylor et al. 1997) 
after the {\em brightness maximum phase}, the 
attention is immediately drawn to the fact that  the  decrease  of 
the ``twinkling" amplitude of the point-like radio source  begins
approximately from the 35th day. The radio fluxes in 8.46 and  4.86
GHz VLA bands show the tendency to stabilization. 
The stabilization of the radio fluxes  began  approximately 
at the same time of the deceleration fading we see in the  $ I_{c} $ filter
($\approx  8000$ \AA).
The ``radio plateau" was lasting 
till about the 65th day after the GRB. Then the flux decreased at both 
frequencies to $\sim 0.4$ mJy by the 87th day. (The  flux  in  the
1.43 GHz band also approached the same value.)
Thus, in spite of  sharp  flux  variations  in  the  beginning  of 
observations, about one month after the burst the  radio  source 
considerably calms. 

We note here that HST/NIMCOS flux measurements (Pian et al. 1998) 
in $ H $ band  gave 
$6.2\pm1.5$ $\mu$Jy, and not $3\pm1$ $\mu$Jy, as was expected  for 
power-law brightness fading in ground-based $ I $ and $ K $  bands. 
It was necessary to extrapolate these data to the $ H  $  band 
assuming the same power-law spectrum slope which was observed in the 
brightness maximum phase 
(Metzger  et  al. 1997a; Sokolov et al. 1997). 
We note that the date of the HST observation and 
this deviation from the expected averaged 
power law in the HST/NIMCOS $ H $ band 
are close to the beginning of the deviation from the power law decay 
in our $ I_{c} $ band (Fig. 4) 
and to the ``twinkling" amplitude decreasing date of 
the point-like radio source.

Pedersen et al. (1998) were the first to report about the deviation 
from the average power law. These authors find $R=24.28\pm 0.10$, 96 days
after the high energy event,
which gives a deviation from the power law decay of $\approx 0.6$ mag. From the
HST observation (on June 2, 1997) we see that the object  
retained the point-like appearance with a possible 
contribution of an underlying ``compact galaxy" with FWHM = $0''.15$ 
and with $R\ge  25.4$. 
The possible deceleration of brightness fading rate in HST/NIMCOS $ H $ band
and the observed flattening of the light curve in the $I_{c}$ filter 
36 days after the GRB
might be interpreted as a sign of the detection of such a galaxy. 
(This would imply a quite red object for $ z = 0.835 $, 
but we do not discuss this implication in this work.) 

On the other hand, the multi-wavelenght (optical, X-ray, radio) 
behavior we observe can also be explained by 
some intrinsic features of GRB970508 OT, 
rather than by the manifestation of an host galaxy. 
For example, it is possible to suppose that we observe the formation 
of a jet-like structure related to the activity of a 
stellar mass compact object, i.e. a {\em gamma-ray burster}. 
The possibility of explanation of all the observational data by 
the intrinsic features of the source
in the model of asymmetrical relativistic jets is now actively
discussed (Dar 1997; Paczyn'ski 1997). 

\section{Summary}

The variable optical source associated to GRB970508 was observed
with the 1-m and 6-m telescopes of SAO RAS from May 9  
to August 4, 1997. The details of temporal changes of brightness in the
$BVR_{c}I_{c}$ bands were compared with observations in X-ray and radio bands.
We confirm the burst activity of the object detected by Piro et al. (1998)
in X-rays in the first week after the high energy event. We notice also
that the further brightness fading of the point-like object 
deviates from the average power law. 
This behavior was clearly seen at
$\approx 8000$ \AA\ when, approximately 36 days after the GRB, a period of
constant brightness starts.

Taking into account the global multi-wavelength behavior
of GRBR 970508, it is presumable that the burst activity and
subsequent deceleration of the brightness fading of the optical
stellar-like source is caused by some intrinsic features of the object itself. 
Alternatively we probably observe the displaying of a dwarf underlying host
galaxy with $I_c \approx 23.1$. 
To be certain of the absence of appreciable variability of 
a conjectural constant source further observational monitoring 
of a ``quiet state" of the GRBR 970508 could give crucial inputs 
to the solution of this mystery.

{\em Acknowledgements}: We thank: the BeppoSAX team for the quick alert of the
 GRB970508 event; the 6-m Telescope Program Committee for 
allocating the observational time for the program of gamma-ray bursters 
identification; the Director of SAO Yu.Yu. Balega for his support to our 
program; 
the observers Yu.N. Parijskij, V.H. Chavushyan, N.A. Tikhonov, L.I. Snezhko
and T.N. Sokolova for her help during the preparation of this text. \\
The work was carried out with the support of the RF {\em Astronomy} foundation 
(grant 97/1.2.6.4) and also INTAS N96-0315.

\end{document}